\documentclass[12pt]{article}
\pdfoutput=1
\usepackage[english]{babel}
\usepackage[latin1]{inputenc}
\usepackage{epsfig}
\usepackage{color}
\usepackage{rotating}
\usepackage{psfrag}
\usepackage{epstopdf}

\usepackage[intlimits]{amsmath}
\usepackage{amssymb}
\usepackage{slashed}
\usepackage{feynmp}

\usepackage{url}
\usepackage{graphicx}

\numberwithin{equation}{section}

\long\def\symbolfootnote[#1]#2{\begingroup\def\thefootnote{\fnsymbol{footnote}}
\footnote[#1]{#2}\endgroup}

\newcommand{\GeV}{\,{\rm GeV}}

\newcommand{\TeV}{\,{\rm TeV}}
\newcommand{\s}{\,{\rm s}}
\newcommand{\sr}{\,{\rm sr}}
\newcommand{\cm}{\,{\rm cm}}

\newcommand{\kpc}{\,{\rm kpc}}

\begin{document}

\setlength{\unitlength}{1mm}

\date{\mbox{ }}

\title{
{\normalsize
TUM-HEP 839/12\hfill\mbox{}\\}
\vspace{1.5cm} 
\bf Probing the stability of superheavy dark matter particles with high-energy neutrinos\\[8mm]}

\author{Arman Esmaili$^1$, Alejandro Ibarra$^2$, Orlando L. G. Peres$^{1,3}$ \\[2mm]
{\normalsize\it $^1$Instituto de F\'isica Gleb Wataghin - UNICAMP, 13083-859, Campinas, SP, Brazil}\\[-0.05cm]
{\normalsize\it $^2$Physik-Department T30d, Technische Universit\"at M\"unchen,}\\[-0.05cm]
{\it\normalsize James-Franck-Stra\ss{}e, 85748 Garching, Germany}\\[-0.05cm]
{\normalsize\it $^3$The Abdus Salam International Centre for Theoretical Physics, I-34100 Trieste, Italy}}

\maketitle

\thispagestyle{empty}

\begin{abstract}
\noindent 
Two of the most fundamental properties of the dark matter particle, the mass and the lifetime, are only weakly constrained by the astronomical and cosmological evidence of dark matter. We derive in this paper lower limits on the lifetime of dark matter particles with masses in the range $10~{\rm TeV}-10^{15}~{\rm TeV}$ from the non-observation of ultrahigh energy neutrinos in the AMANDA, IceCube, Auger and ANITA experiments. For dark matter particles which produce neutrinos in a two body or a three body leptonic decay, we find that the dark matter lifetime must be longer than $\mathcal{O}(10^{26}-10^{28})~{\rm s}$ for masses between 10 TeV and the Grand Unification scale. Finally, we also calculate, for concrete particle physics scenarios, the limits on the strength of the interactions that induce the dark matter decay.
\end{abstract}

\newpage

\section{Introduction}

There is currently mounting evidence for the existence of dark matter (DM) in our Universe from various astronomical and cosmological observations~\cite{Bertone:2004pz}. However, very fundamental properties of the dark matter particle such as the mass or the lifetime are still fairly unconstrained. An important restriction on dark matter models arises from large-scale $N$-body simulations~\cite{Navarro:1995iw,Springel:2005nw}  which indicate that, in order to reproduce the observations of large scale galaxy surveys, the dark matter particle moved very non-relativistically at the time of structure formation, namely the dark matter particle is ``cold''. For dark matter particles which reached thermal equilibrium with the primordial plasma, this condition is achieved when the dark matter mass is $m_{\rm DM}\gtrsim 1$ keV~\cite{Viel:2005qj}, provided their relic density today coincides with the dark matter abundance inferred from the seven-year WMAP data, $\Omega_{\rm DM}=0.23$~\cite{Komatsu:2010fb}. Furthermore, for this class of models, the partial wave unitarity of the $S$ matrix implies the upper limit $m_{\rm DM}\lesssim 100 $ TeV~\cite{Griest:1989wd}. In contrast, for particles which were never in thermal equilibrium, the allowed dark matter mass window ranges from $10\, \mu$eV (as for dark matter axions~\cite{Sikivie:2006ni}) to the Grand Unification scale~\cite{Chung:1998zb,Kolb:1998ki,Kolb:2007vd}.

Besides, the existing evidence for dark matter does not require the dark matter particle to be absolutely stable, as commonly assumed in the literature. Namely, whereas the observed longevity of the dark matter particle can be attributed to a symmetry which is approximately conserved, this symmetry could be broken by physics at very high energies, thus inducing the dark matter decay. This rationale is completely analogous to the proton stability, which can be attributed to a baryon number symmetry but which could nonetheless be broken by new physics at the grand unification scale, in turn inducing the proton decay. Conversely, the study of the dark matter stability provides valuable information about the dark matter interactions with the Standard Model particles and about the symmetries that ensure the longevity of the dark matter particle.

In this paper we will derive limits on the dark matter lifetime focusing on the case where the dark matter particle is superheavy, complementing recent analyses studying the stability of dark matter particles with masses in the range 1 GeV-10 TeV \cite{Ibarra:2008jk,Meade:2009iu,Ibarra:2009dr,Chen:2009uq,PalomaresRuiz:2007ry,Covi:2009xn,Zhang:2009ut,Vertongen:2011mu,Garny:2010eg,Huang:2011xr}. The decay of superheavy dark matter particles has been considered in the past as a source of high energy cosmic rays \cite{Frampton:1980rs,Ellis:1980ap,Hill:1982iq,Ellis:1990nb,Berezinsky:1997hy,Sarkar:2001se,Barbot:2002ep,Aloisio:2003xj}. In the so-called ``top-down'' models, the decay of the dark matter particles into partons produces a flux of cosmic protons, which is accompanied by a flux of gamma-rays and neutrinos. Neutrinos have the property that they travel in the Universe without suffering an appreciable attenuation, making them a very suitable messenger to constrain the lifetime of superheavy dark matter particles; the possibility of detecting of neutrinos in ``top-down'' models has been discussed in \cite{Sarkar:2001se,Barbot:2002ep,Aloisio:2003xj,Barbot:2002kh}. In this paper we will concentrate on scenarios where the dark matter particle decays leptonically, producing a flux of high energy neutrinos. To derive bounds on the lifetime we will use the limits recently published by various experiments on the high energy neutrino flux and which cover a wide range of energies: AMANDA between $16\,{\rm TeV}$ and $2.5\times 10^3\,{\rm TeV}$~\cite{Achterberg:2007qp}, IceCube between $340\,{\rm TeV}$ and $6.3\times 10^6\,{\rm TeV}$~\cite{Abbasi:2012cu,Abbasi:2011ji}, Auger between $10^5\,{\rm TeV}$ and $10^8\,{\rm TeV}$ ~\cite{Abreu:2011zze} and ANITA between $10^6\,{\rm TeV}$ and $3.2\times 10^{11}\,{\rm TeV}$~\cite{Gorham:2010kv}. All these experiments probe different energy windows and perfectly complement each other, allowing us to derive stringent limits on the lifetime of dark matter particles with masses ranging between 10 TeV and the Grand Unification scale.

The paper is organized as follows: in sec.~\ref{sec:flux} we discuss the expected neutrino flux from both two body and three body decays of dark matter particles, taking into account the opacity of the Universe to ultra high-energy neutrinos. Using the expected neutrino flux, we derive in sec.~\ref{sec:limit} lower limits on the dark matter lifetime from the data obtained by the AMANDA, IceCube, Auger and ANITA experiments. In sec.~\ref{sec:models} we translate the derived limits on the dark matter lifetime into limits on the coupling constants of effective Lagrangians describing concrete scenarios of dark matter decay.  Lastly, in sec.~\ref{sec:conc} we present our conclusions.

\section{\label{sec:flux}Neutrino flux from dark matter decay}

The neutrino flux from dark matter decay receives two contributions.
The first one stems from the decay of dark matter particles in
the Milky Way halo and leads to a differential flux given by:
\begin{equation}
  \frac{dJ_\text{halo}}{dE_\nu}(l,b) = 
  \frac{1}{4\pi\,m_{\rm DM}\,\tau_{\rm DM}} 
  \frac{dN_\nu}{dE_\nu}
  \int_0^\infty ds\; 
  \rho_{\rm halo}[r(s,l,b)] \;,
  \label{halo-flux}
\end{equation}
where $\rho_{\rm halo}(r)$ is the density profile of dark matter particles in our Galaxy as a function of the distance from the Galactic center, $r$, and $dN_\nu/dE_\nu$ is the energy spectrum of neutrinos produced in the decay of a dark matter particle. The neutrino flux received at Earth depends on the Galactic coordinates, longitude $l$ and latitude $b$, and is given by a line-of-sight integral over the parameter $s$, which is related to $r$ by
\begin{equation}\label{galcoord}
  r(s,l,b) = \sqrt{s^2+R^2_\odot-2 s R_\odot \cos b\cos l}\,,
\end{equation}
where $R_\odot=8.5\kpc$ is the distance of the Sun to the Galactic center.

The average flux over the full sky can then be straightforwardly calculated, the result being
\begin{equation}\label{eq:halo}
\frac{dJ_{\rm halo}}{dE_\nu}=  D_{\rm halo}  \frac{dN_\nu}{dE_\nu}\;.
\end{equation}
For our numerical analysis we will adopt a Navarro-Frenk-White 
density profile~\cite{Navarro:1996gj}
\begin{equation}
\rho_\text{halo}(r)\simeq \frac{\rho_h}{r/r_c (1+r/r_c)^2}\;,
\end{equation}
where $r$ is the distance to the Galactic center, $r_c\simeq20\,\text{kpc}$ is the critical radius and $\rho_h\simeq0.33\GeV \cm^{-3}$, which yields a dark matter density at the Solar System $\rho_\odot~=~0.39\GeV \cm^{-3}$~\cite{Catena:2009mf}. For this choice of parameters,
\begin{equation}\label{eq:Dhalo}
D_{\rm halo}=1.7\times 10^{-8} 
\left(\frac{1 \TeV}{m_{\rm DM}}\right)
\left(\frac{10^{26}\s}{\tau_{\rm DM}}\right)\cm^{-2}\s^{-1}\sr^{-1}\;.
\end{equation}

In addition to the neutrino flux from the decay of dark matter particles in the Milky Way halo, there is a second contribution stemming from the decay of dark matter particles at cosmological distances, which produces a perfectly isotropic diffuse neutrino flux. The differential flux with respect to the received neutrino energy is given by:
\begin{equation}
  \begin{split}
    &{}\frac{dJ_\text{eg}}{dE_\nu} =
    \frac{\Omega_{\rm DM}\rho_\text{c}}{4\pi m_{\rm DM} \tau_{\rm DM}}
    \int_0^\infty dz
    \frac{1}{H(z)}
    \;\frac{dN_\nu}{dE_\nu}\left[(1+z)E_\nu\right]
    \;e^{-s_\nu(E_\nu,z)}\;,
  \end{split}
  \label{eqn:NuFluxEG}
\end{equation}
where $H(z)=H_0 \sqrt{\Omega_\Lambda+\Omega_\text{m}(1+z)^3}$ is the Hubble expansion rate as a function of redshift $z$ and $\rho_\text{c}=5.5\times10^{-6}\GeV/\cm^3$ denotes the critical density of the Universe. Throughout this work we assume a $\Lambda$CDM cosmology with parameters $\Omega_\Lambda=0.73$, $\Omega_\text{m}=0.27$, $\Omega_{\rm DM}=0.23$ and $h\equiv H_0/100\,\text{km}\;\text{s}^{-1}\,\text{Mpc}^{-1}=0.70$, as derived from the seven-year WMAP data~\cite{Komatsu:2010fb}. Lastly, $s_\nu(E_\nu,z)$ is the neutrino opacity of the Universe, which was calculated in \cite{Gondolo:1991rn,Weiler:1982qy}, assuming that the neutrinos are massless, for an age of the Universe $t_0=6.5$ Gyr. With the most recent determination of the age of the Universe, $t_0=13.76$ Gyr~\cite{Komatsu:2010fb}, the neutrino opacity reads:
\begin{equation}
s_\nu(E_\nu,z)=\begin{cases}
7.4 \times 10^{-17} (1+z)^{7/2} (E_\nu/\TeV), & {\rm for}~1\ll z < z_{\rm eq} \\
1.7 \times 10^{-14} (1+z)^3 (E_\nu/\TeV), & {\rm for}~ z\gg z_{\rm eq} \\
\end{cases}
\end{equation}
where $z_{\rm eq}\sim 10^4$ is the redshift of the equality of matter and radiation in the Universe. 

In this paper we will analyze the following scenarios. First, we will study the limits on the lifetime of a scalar dark matter particle which decays into a neutrino and an antineutrino, $\phi_{\rm DM}\rightarrow \nu \bar \nu$. In this case, the energy spectrum reads:
\begin{equation}
\frac{dN_\nu}{dE_\nu}=\delta\left(E_\nu-\frac{m_{\rm DM}}{2}\right)\;.
\label{eq:dNdE2body}
\end{equation}
Our second scenario consists of a Majorana dark matter particle which decays into a photon and a (anti)neutrino, $\psi\rightarrow \gamma \nu (\bar\nu)$. Due to the Majorana nature of the dark matter particle, the branching ratio for each decay channel is, at tree level, equal to 50\%. In this scenario the energy spectrum of the (anti)neutrinos is also given by Eq.~(\ref{eq:dNdE2body}) and displays a very sharp line feature. To evaluate the limits on models which produce neutrinos with a softer spectrum we will also study a scenario consisting of dark matter particles which decay into an electron-positron pair and a (anti)neutrino $\psi\rightarrow e^+e^- \nu(\bar\nu)$. 
As before, the branching fraction for each decay channel is $\sim 50\%$. In this case the energy spectrum of the neutrinos is fairly model dependent, namely it depends on whether the decay is mediated by a heavy charged scalar or a heavy charged vector and on the chirality of the Standard Model fermions to which the dark matter particle couples~\cite{Garny:2010eg}. When the dark matter particle couples just to the left-handed electrons we find
\begin{equation}
\frac{dN_\nu}{dE_\nu}=\frac{2}{E_\nu}\left(\frac{2E_\nu}{m_{\rm DM}}\right)^3
\left[3-2 \left(\frac{2E_\nu}{m_{\rm DM}}\right)\right]~,
\label{eq:dNdE3body}
\end{equation}
both in the case when the decay is mediated by a scalar or by a vector.

\begin{figure}[t]
\begin{center}
\includegraphics[width=0.7\textwidth]{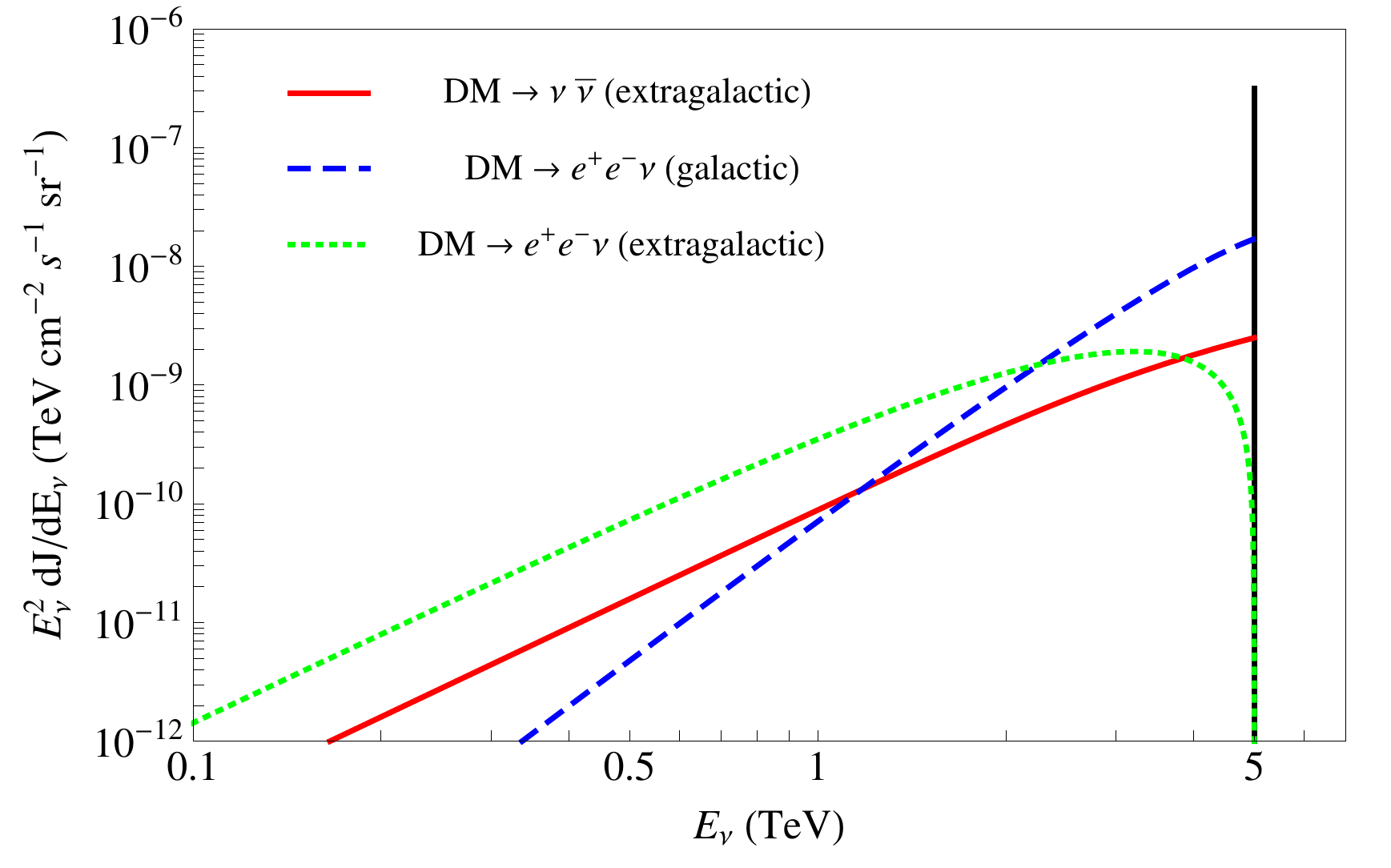}
\caption{Contributions to the neutrino flux at the Earth from two body and three body decay of a dark matter particle with $m_{\rm DM}=10$~TeV and $\tau_{\rm DM}=10^{26}$~s.  }
\label{fig:spectrum}
\end{center}
\end{figure}

We show in Fig.~\ref{fig:spectrum} the various contributions to the differential neutrino flux for the two body and for the three body dark matter decays, Eqs.~(\ref{eq:dNdE2body},\ref{eq:dNdE3body}) respectively. In this figure we assumed $m_{\rm DM}=10$~TeV and $\tau_{\rm DM}=10^{26}$~s. As can be seen from Fig.~\ref{fig:spectrum}, the maximum energy of the neutrino is $m_{\rm DM}/2$ and the vertical black solid line represents the sharp line in the neutrino spectrum coming from the two body decay $\phi\to\nu\bar{\nu}$ of dark matter particles in the galactic halo. The red solid line shows the extragalactic contribution for the two body dark matter decay (see Eq.~(\ref{eqn:NuFluxEG})). The dashed (blue) and dotted (green) curves show respectively the galactic and extragalactic contributions for the three body decay $\psi\to e^+ e^- \nu$. It is apparent from the plot that the neutrino flux at the energies near to $m_{\rm DM}/2$ is dominated by the halo contributions while at lower energies by the extragalactic contribution. This feature exists for all the dark matter masses and, as we will see in the next section, will play a role in extracting the lower limit on the lifetime of superheavy dark matter particles.

The initial flavor composition of the neutrino flux from decaying dark matter at the production point $J^0_e:J^0_\mu:J^0_\tau$ depends on the details of the dark matter model. After production, neutrinos travel long distances to the Earth (of the order $\sim$~kpc for the Galactic decaying dark matter particles and $\gtrsim$~Mpc for the extragalactic ones). Due to these long distances, the neutrinos arrive at the Earth decoherently, such that the flavor composition of the neutrino flux at the Earth is $(J^\oplus_e,J^\oplus_\mu,J^\oplus_\tau)^T=\mathbb{P}\,(J^0_e,J^0_\mu,J^0_\tau)^T$, where the elements of the symmetric oscillation probability matrix $\mathbb{P}$ are given by $P_{\alpha\beta}=\sum_i |U_{\alpha i}|^2|U_{\beta i}|^2$. Assuming the best-fit values for the mixing parameters~\cite{Tortola:2012te} including the updated result on $\theta_{13}$ from the recent measurements ($\sin^2 2\theta_{13}=0.1$~\cite{Ahn:2012nd}), the symmetric matrix $\mathbb{P}$ is given by 

\begin{equation}
\mathbb{P}=\left( \begin{matrix}
      P_{ee} & P_{e\mu} & P_{e\tau} \\
      & P_{\mu\mu} & P_{\mu\tau} \\
      & & P_{\tau\tau} \\
   \end{matrix} \right) = \left( \begin{matrix}
      0.54 & 0.26 & 0.2 \\
      & 0.37 & 0.37 \\
      & & 0.42 \\
       \end{matrix} \right)~.
\end{equation}   

Comparing the values of the elements of matrix $\mathbb{P}$, it follows that even if the dark matter decay produces neutrinos with just one flavor, similar fluxes are received at the Earth for all the three flavors. Therefore, in our numerical results we will assume for simplicity that the flavor composition of the neutrinos are in the ratios $J^\oplus_e:J^\oplus_\mu:J^\oplus_\tau=1:1:1$. Any deviation from this assumption leads to at most a factor of two difference in our limits.

\section{\label{sec:limit}Limits on the dark matter lifetime}

In this section we will present the limits on the dark matter lifetime from various experiments including AMANDA~\cite{Achterberg:2007qp}, IceCube-22~\cite{Abbasi:2012cu}, IceCube-40~\cite{Abbasi:2011ji}, Auger~\cite{Abreu:2011zze} and ANITA~\cite{Gorham:2010kv}. However before discussing each experiment, let us  first describe the method that was employed to derive the limits. Generally, for each of these experiments, it is possible to define an ``effective area'' $A^\alpha_{\rm eff}$ for the detection of $\nu_\alpha+\bar{\nu}_\alpha$ ($\alpha=e,\mu,\tau$). By definition, the effective area is the hypothetical area of the experiment with 100\% efficiency which can be obtained after developing appropriate cuts for the background-signal discrimination. The expected number of events is given by
\begin{equation}\label{eq:Nexp}
N_{\rm exp}=T \Delta\Omega \sum_\alpha\left[ \int_{E^{\rm min}_{\nu}}^{E^{\rm max}_{\nu}} \frac{{\rm d}J^\oplus_{\nu_\alpha+\bar{\nu}_\alpha}}{{\rm d}E_{\nu_\alpha,\bar{\nu}_\alpha}} A_{\rm eff}^\alpha (E_{\nu}) {\rm d}E_{\nu}\right]~,
\end{equation}
where ${\rm d}J^\oplus_{\nu_\alpha+\bar{\nu}_\alpha}/{\rm d}E_{\nu}$ is the energy spectrum of $\nu_\alpha+\bar{\nu}_\alpha$ flux at the surface of Earth, $T$ is the live time of the experiment, $\Delta\Omega$ is the solid angle acceptance of the experiment, and $E^{\rm min}_{\nu}$ and $E^{\rm max}_{\nu}$ are, respectively, the minimum and maximum of the neutrino energy considered in the data analysis. The upper limit on the number of neutrinos from the diffuse flux depends on the number of signal candidates $N_{\rm sig}$ (after performing appropriate cuts) and the expected number of background events $N_{\rm bg}$. For a given $N_{\rm sig}$ and $N_{\rm bg}$, the Bayesian upper limit $N_{\rm limit}$ on $N_{\rm exp}$ at $q\%$ confidence level can be extracted from the following equation 
\begin{equation}\label{eq:bayesian}
q/100=\frac{\int_0^{N_{\rm limit}} L(N_{\rm sig}|N)\, {\rm d}N}{\int_0^\infty L(N_{\rm sig}|N)\, {\rm d}N}~,
\end{equation}  
where $L(N_{\rm sig}|N)$ is the likelihood function for Poisson distributed variable $N_{\rm sig}$ which is given by
\begin{equation}
L(N_{\rm sig}|N)=\frac{(N+N_{\rm bg})^{N_{\rm sig}}}{N_{\rm sig}!} \,e^{-(N+N_{\rm bg})}~.
\end{equation}
After extracting the value of $N_{\rm limit}$ from Eq.~(\ref{eq:bayesian}), the upper limit on the diffuse flux of neutrinos at a certain confidence level can be derived from $N_{\rm exp}\leq N_{\rm limit}$~.

Various experiments have recently searched for a diffuse flux of ultra-high energy neutrinos. The AMANDA experiment has searched for a diffuse flux of muon neutrinos in the energy range $16~{\rm TeV}<E_\nu<2.5\times10^{3}~{\rm TeV}$, finding $N_{\rm sig}=7$ signal events while the expected number of background events is $N_{\rm bg}=6$, hence giving a 90\% C.L. limit on the total number of neutrinos in that energy window $N_{\rm limit}=5.4$~\cite{Achterberg:2007qp}. Besides, the IceCube collaboration has undertaken a dedicated analysis searching for ultra-high energy tau neutrinos with the 22-strings configuration, without discriminating between electron and muon neutrinos~\cite{Abbasi:2012cu}. The energy range of the analysis is $340~{\rm TeV}<E_\nu<2\times 10^5~{\rm TeV}$ and the expected number of background events is $N_{\rm bg}=0.6$. In this analysis the number of candidates of signal events is $N_{\rm sig}=3$, which results in $N_{\rm limit}=6.1$ at 90\% C.L. An extended analysis using the 40-strings configuration yielded $N_{\rm sig}=0$ in the energy range $2\times10^3~{\rm TeV}<E_\nu<6.3\times10^6~{\rm TeV}$ while $N_{\rm bg}=0.1$~\cite{Abbasi:2011ji}, which translates into  $N_{\rm limit}=2.3$. Furthermore, the Auger experiment has covered an energy window with extends to higher energies, $10^5~{\rm TeV}<E_\nu<10^8~{\rm TeV}$. In this case, $N_{\rm bg}=0$ and $N_{\rm sig}=0$~\cite{Abreu:2011zze}, which results in $N_{\rm limit}=2.3$. Lastly, in the extremely high energy region, the ANITA experiment has covered the window $10^6~{\rm TeV}<E_\nu<3.2\times 10^{11}~{\rm TeV}$, finding $N_{\rm sig}=1$ while $N_{\rm bg}=0.97$~\cite{Gorham:2010kv}, giving $N_{\rm limit}=3.3$. Table~\ref{table:analysis} summarizes the energy window covered by each experiment and the relevant experimental results for our analysis. Also, using the information in Table~\ref{table:analysis}, it is possible to obtain the upper limit on the astrophysical diffuse flux of neutrinos. Assuming the spectrum $k E_\nu^{-2}$ for the diffuse flux, we obtained upper limits $(7.4\times 10^{-8},1.6\times 10^{-7},3.6\times 10^{-8},1.7\times 10^{-7},1.3\times 10^{-7})$~GeV~cm$^{-2}$~s$^{-1}$~sr$^{-1}$ on the normalization factor $k$, respectively for the experiments AMANDA, IC-22, IC-40, Auger and ANITA; which matched the reported limits by each experiment.   

\begin{table}[h]
\caption{Summary of the experiments used in deriving the limit on dark matter lifetime, with their corresponding energy window of sensitivity, the number of background events, $N_{\rm bg}$, the number of signal events, $N_{\rm sig}$, and the upper limit on neutrino events at 90\% C.L., $N_{\rm limit}$.}

\begin{center}
\begin{tabular}{|c|c|c|c|c|}
\hline
 & $E_\nu^{\rm min}-E_\nu^{\rm max}$ (TeV) & $N_{\rm bg}$ & $N_{\rm sig}$ & $N_{\rm limit}$\\
 \hline
AMANDA & $16-2.5\times 10^3$ & 6 & 7 & 5.4  \\
\hline
IceCube-22 & $340-2\times 10^5$ & 0.6 & 3 & 6.1  \\
\hline
IceCube-40 & $2\times 10^3-6.3\times 10^6$ & 0.1 & 0 & 2.3  \\
\hline
Auger & $10^5-10^8$ & 0 & 0 & 2.3  \\
\hline
ANITA & $10^6-3.2\times 10^{11}$ & 0.97 & 1 & 3.3  \\
\hline
\end{tabular}
\end{center}
\label{table:analysis}
\end{table}%

\begin{figure}[h!]
\begin{center}
\includegraphics[width=1.\textwidth]{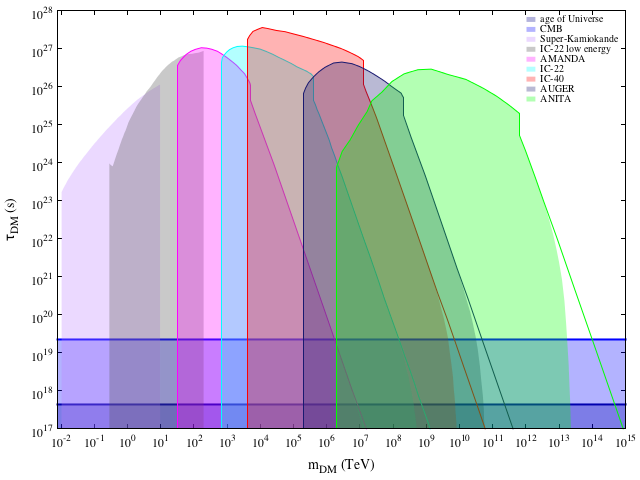}
\caption{Limits (90\% C. L.) on the dark matter lifetime $\tau_{\rm DM}$ from various experiments, assuming two body decay $\phi\to \nu\bar{\nu}$. The horizontal lines at $4.3\times 10^{17}$~s and $2.2\times 10^{19}$~s show respectively the age of Universe and the limit derived in \cite{Gong:2008gi} from the position of the first peak of the cosmic microwave background. The excluded regions from left to right respectively correspond to the experiments: Super-Kamiokande taken from~\cite{Covi:2009xn}, IceCube-22 low energy taken from~\cite{Abbasi:2011eq}, AMANDA, IceCube-22, IceCube-40, Auger and ANITA.}
\label{fig:twobodylimit}
\end{center}
\end{figure}

\begin{figure}[h!]
\begin{center}
\includegraphics[width=1.\textwidth]{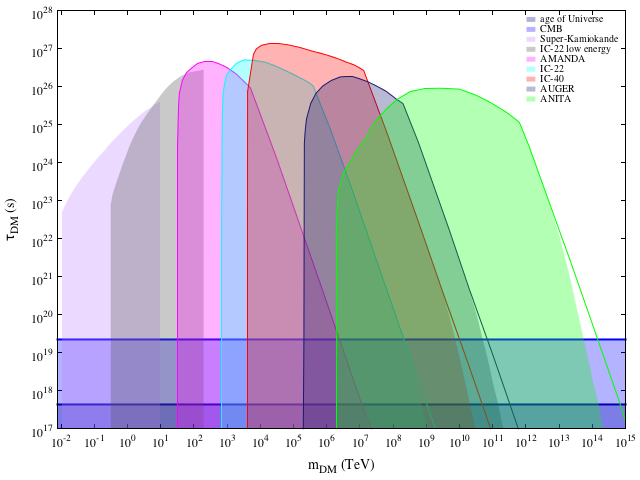}
\caption{Same as Figure \ref{fig:twobodylimit}, assuming three body decay $\psi\to e^+ e^- \nu$.}
\label{fig:threebodylimit}
\end{center}
\end{figure}

For each of these experiments, we derive 90\% C.L. limits on the dark matter lifetime as a function of the dark matter mass requiring that the number of expected events from dark matter decay does not exceed $N_{\rm limit}$, {\it cf.} Table~\ref{table:analysis}. To this end, we use the ``effective area'' reported by each of these experiments and calculate the total number of events from Eq.(\ref{eq:Nexp}), taking the neutrino fluxes in Eqs.~(\ref{eq:halo},\ref{eqn:NuFluxEG}) for the two and the three body decay spectra, Eqs.~(\ref{eq:dNdE2body},\ref{eq:dNdE3body}). Fig.~\ref{fig:twobodylimit} shows the 90\% C. L. limit on $\tau_{\rm DM}$ for the scenario where the dark matter particle is a scalar which decays into a neutrino and an antineutrino, $\phi\rightarrow \nu\bar \nu$. The horizontal lines at $4.3\times 10^{17}$~s and $2.2\times 10^{19}$~s show respectively the age of Universe and the limit derived from the position of the first peak of the cosmic microwave background~\cite{Gong:2008gi}. The excluded regions from left to right correspond, respectively, to the experiments Super-Kamiokande, IceCube-22 low energy, AMANDA, IceCube-22, IceCube-40, Auger and ANITA. The Super-Kamiokande limit was taken from~\cite{Covi:2009xn}, while the ``IC-22 low energy" excluded gray region from~\cite{Abbasi:2011eq}, which performed a dedicated search for the two body decay of dark matter particles in the galactic halo employing the IC-22 configuration; all other limits are new, to the best of our knowledge. The limits derived for each experiment consist of two parts: 1) a peak-shaped region which stems from the galactic halo contribution to the neutrino flux and which corresponds to dark matter masses in the range $E_\nu^{\rm min}\leq m_{\rm DM} /2 \leq E_\nu^{\rm max}$, namely to dark matter scenarios where the produced neutrino has an energy falling in the window to which the experiment is sensitive; 2) a high-energy tail, which corresponds to dark matter masses $ m_{\rm DM} /2 \geq E_\nu^{\rm max}$, such that only the neutrinos produced in decays at large redshifts have energies in the window of sensitivity of the experiment. For each experiment in Fig.~\ref{fig:twobodylimit}, the solid line in the tail part shows the limit neglecting the opacity of the Universe, $s_\nu$, while the shaded area shows the effect of the opacity. As apparent from the plot, the absorption of neutrinos only plays a role in the extremely high energy range and effectively restricts the capabilities of experiments to limit the lifetime of dark matter particles with mass around the Planck scale.

On the other hand, Fig.~\ref{fig:threebodylimit} shows the 90\% C. L. limit on the dark matter lifetime $\tau_{\rm DM}$ assuming the three body decay $\psi\to e^+ e^- \nu(\bar{\nu})$. It is noticeable that, due to the larger extragalactic flux at low energies for heavy dark matter particles, the limits from the tail part in Fig.~\ref{fig:threebodylimit} are stronger than the corresponding limits in Fig.~\ref{fig:twobodylimit}. Furthermore, for a given experiment, the limits in Fig.~\ref{fig:threebodylimit} for dark matter masses in the range $E_\nu^{\rm min}\leq m_{\rm DM} /2 \leq E_\nu^{\rm max}$  are approximately a factor of two smaller than those in Fig.~\ref{fig:twobodylimit}. This can be easily understood calculating analytically, from Eq.~(\ref{eq:Nexp}), the lower limits on the two body decay and three body decay lifetimes, $\tau^{\rm 2-body}_{\rm limit}$ and $\tau^{\rm 3-body}_{\rm limit}$. Taking into account that the neutrinos from the two-body decay of dark matter particles in the halo are monoenergetic and neglecting the contribution from the cosmological decays, it is straightforward to calculate $\tau^{\rm 2-body}_{\rm limit}$ as the values of $\tau_{\rm DM}$ which saturate the following inequality
\begin{equation}\label{eq:2-3relation1}
N_{\rm exp}\simeq T \Delta\Omega \,\frac{1}{m_{\rm DM} \tau_{\rm DM}}\,\frac{d}{3}\,\sum_\alpha A_{\rm eff}^\alpha \left(\frac{m_{\rm DM}}{2}\right)\leq N_{\rm limit}~,
\end{equation}
where $d$ is the numerical factor in Eq.~(\ref{eq:Dhalo}), which is equal to $1.7\times 10^{-8}\cm^{-2}\s^{-1}\sr^{-1}$ for $m_{\rm DM}=1\,{\rm TeV}$ and $\tau_{\rm DM}=10^{26}\,{\rm s}$, and the factor $1/3$ comes from the assumption of democratic flavor composition of neutrinos at the Earth $J^\oplus_e:J^\oplus_\mu:J^\oplus_\tau=1:1:1$. Conversely, using  Eq.~(\ref{eq:2-3relation1}) to cast the effective area in terms of $\tau^{\rm 2-body}_{\rm limit}(m_{\rm DM})$, it is easy to show that the limit on the three body decay lifetime $\tau^{\rm 3-body}_{\rm limit}$ can be written as
\begin{equation}\label{eq:2-3relation2}
\tau^{\rm 3-body}_{\rm limit}(m_{\rm DM})\simeq\frac{1}{m_{\rm DM}}\frac{1}{2} \int_{E^{\rm min}_\nu}^{E^{\rm max}_\nu} \frac{{\rm d}N_\nu}{{\rm d}E_\nu} \,(2E_\nu)\, \tau^{\rm 2-body}_{\rm limit}(2E_\nu) \,{\rm d}E_\nu~,
\end{equation}
where ${\rm d}N_\nu/{\rm d}E_\nu$ is given in Eq.~(\ref{eq:dNdE3body}) and the factor $1/2$ in front of the integral comes from the fact that the number of neutrinos plus antineutrinos per three body decay is half of the number of neutrinos per two body decay. From Eq.~(\ref{eq:2-3relation2}) it can be checked that, for a given experiment, when $m_{\rm DM}\simeq 2E^{\rm max}_\nu$ the relation $\tau^{\rm 3-body}_{\rm limit}\simeq\tau^{\rm 2-body}_{\rm limit}/2$ approximately holds, as almost all the spectrum of neutrinos lies inside the integration range of the experiment. However, for $m_{\rm DM}<2 E^{\rm max}_\nu$ it follows that $\tau^{\rm 3-body}_{\rm limit}\lesssim \tau^{\rm 2-body}_{\rm limit}/2$ since some part of spectrum lies outside the integration range of the experiment. This procedure has been employed in Fig.~\ref{fig:threebodylimit} to derive the ``IC-22 low energy'' limits from the published lower bounds on the lifetime for the decay $\phi\rightarrow \nu\bar \nu$~\cite{Abbasi:2011eq}; for all other limits, the effective areas provided by the experiments were used.

\section{\label{sec:models}Implications for decaying dark matter models}

\begin{figure}[h!]
\begin{center}
\includegraphics[width=0.8\textwidth]{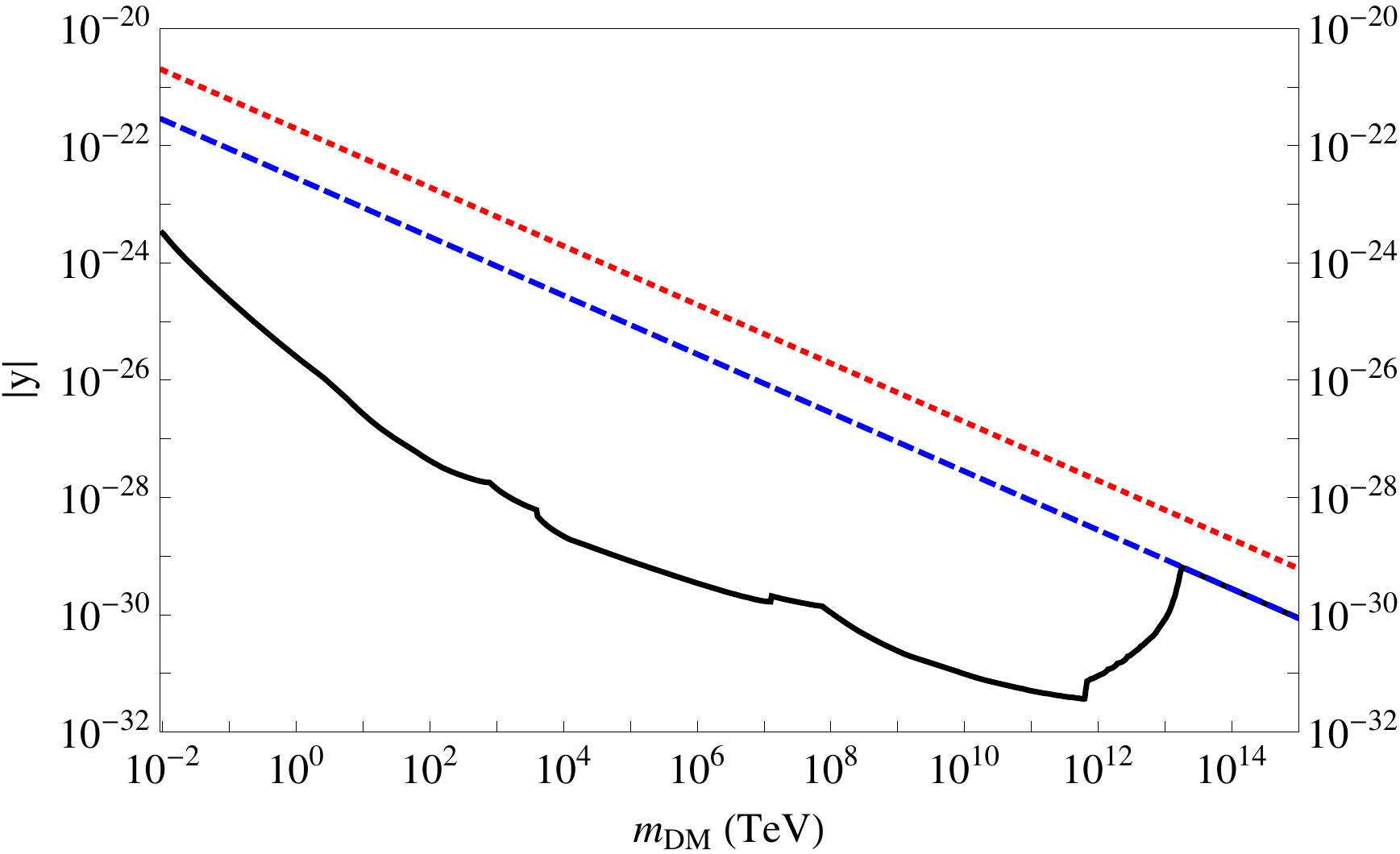}
\caption{ Upper limit (90\% C. L.) on the coupling $y$ which induces the two body decay $\phi\rightarrow \nu\bar \nu$ (see Eq.~(\ref{eq:lambda})). The dotted and dashed lines show respectively the limit from age of Universe and CMB. The solid line shows the limit obtained in this paper. }
\label{fig:lambdalimit}
\end{center}
\end{figure}

In this section we will translate the upper limits on the dark matter lifetime derived in the previous section into limits on the couplings of the effective Lagrangian which induce the dark matter decay. Let us first discuss the scenario where the dark matter particle is a scalar. In this case the effective Lagrangian that induces the decay $\phi_{\rm DM}\rightarrow \nu \bar\nu$ contains a dimension 4 operator:
\begin{equation}
-{\cal L}_{\rm eff}=y \phi_{\rm DM} \bar \nu \nu+{\rm h.c.}
\end{equation}
where $y$ is the coupling constant. The decay rate can be straightforwardly calculated, the result being:
\begin{equation}\label{eq:lambda}
\Gamma(\phi_{\rm DM}\rightarrow \nu \bar\nu)=\frac{|y|^2}{8\pi}m_{\rm DM}~.
\end{equation}
We show in Fig.~\ref{fig:lambdalimit} the upper limits on the coupling $y$ as a function of the dark matter mass from the non-observation of neutrino events in the experiments discussed in sec.~\ref{sec:limit}. The solid line shows the upper limit obtained in this paper, while the dotted line shows the upper limit coming from requiring a dark matter lifetime longer than the age of Universe and the dashed line shows the limit from the CMB data~\cite{Gong:2008gi}.

On the other hand, when the dark matter is a Majorana fermion, the effective Lagrangian that induces the decay $\psi\rightarrow \nu \gamma$ reads~\cite{Giunti:2008ve}
\begin{equation}
{\cal L}=\frac{1}{2}\bar\psi_{\rm DM}\, \sigma_{\alpha\beta}(\mu +\epsilon\gamma_5)\,
\nu F^{\alpha\beta}+{\rm h.c.}
\label{eq:L-fermion-2body}
\end{equation}
where $F^{\alpha\beta}$ is the electromagnetic field strength tensor, while $\mu$ and $\epsilon$ are the magnetic and electric transition moments, respectively. When the dark matter and the neutrino have the same CP parities, the magnetic transition moment vanishes, while when they have oppositte CP parities, the electric transition moment vanishes. In either case, the decay rate can be cast as:
\begin{equation}\label{eq:Gamma-fermion-2body}
\Gamma(\psi_{\rm DM}\rightarrow \nu \gamma)=\frac{|\mu_{\rm eff}|^2}{8\pi} m_{\rm DM}^3\;,
\end{equation}
where we have defined an effective neutrino magnetic moment, $|\mu_{\rm eff}|\equiv\sqrt{|\mu|^2+|\epsilon|^2}$. 

\begin{figure}[t]
\begin{center}
\includegraphics[width=0.8\textwidth]{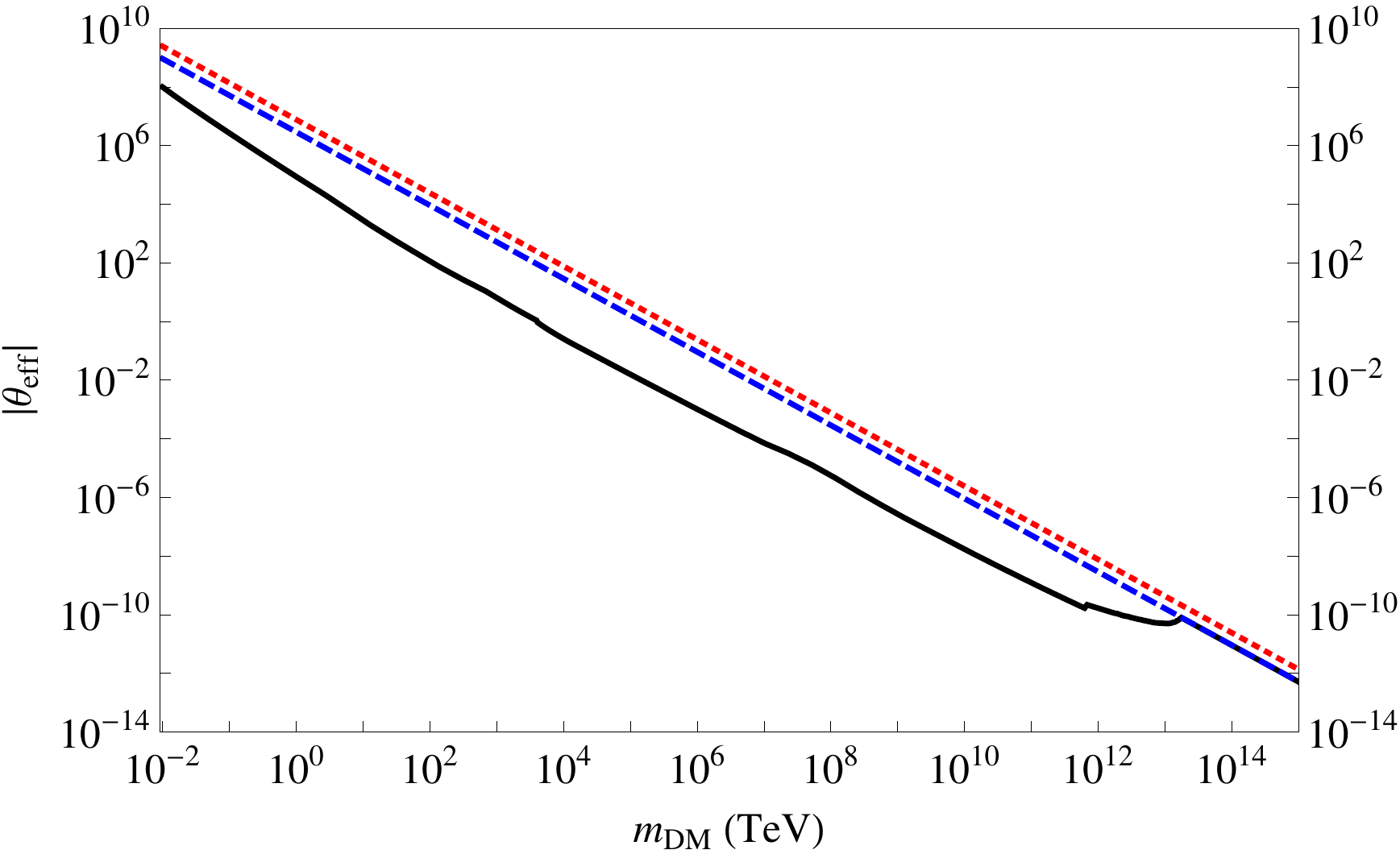}
\caption{Upper limit (90\% C. L.) on the parameter $\theta_{\rm eff}$ which induces the radiative two body decay $\psi\rightarrow \nu \gamma$ (see Eqs.~(\ref{eq:Gamma-fermion-2body},\ref{eq:Mu-fermion-2body})), assuming $M_\Sigma=M_P$. The dotted and dashed lines show respectively the limit from the age of Universe and CMB. The solid line shows the limit obtained in this paper. }
\label{fig:lambdaefflimit}
\end{center}
\end{figure}

In specific models, the effective Lagrangian Eq.(\ref{eq:L-fermion-2body}) is generated by loop effects. Concretely, when the decay is mediated by a heavy scalar circulating in the loop, the effective neutrino magnetic moment can be conveniently parametrized as~\cite{Garny:2010eg}:
\begin{equation}\label{eq:Mu-fermion-2body}
|\mu_{\rm eff}|=\frac{e\, m_{\rm DM}\, |\theta_{\rm eff}|^2}{64\pi^2 M_\Sigma^2}~,
\end{equation}
where $M_\Sigma$ is the mass of the scalar particle in the loop and $\theta_{\rm eff}$ is a combination of the couplings of the dark matter particle and the neutrino to the heavy scalar in the loop. Assuming $M_\Sigma=M_P$, where $M_P$ is the Planck mass, we show in Fig.~\ref{fig:lambdaefflimit} upper limits on the coupling $\theta_{\rm eff}$; when $M_\Sigma<M_P$, the limits scale as $(M_\Sigma/M_P)^4$.

\begin{figure}[h]
\begin{center}
\includegraphics[width=0.8\textwidth]{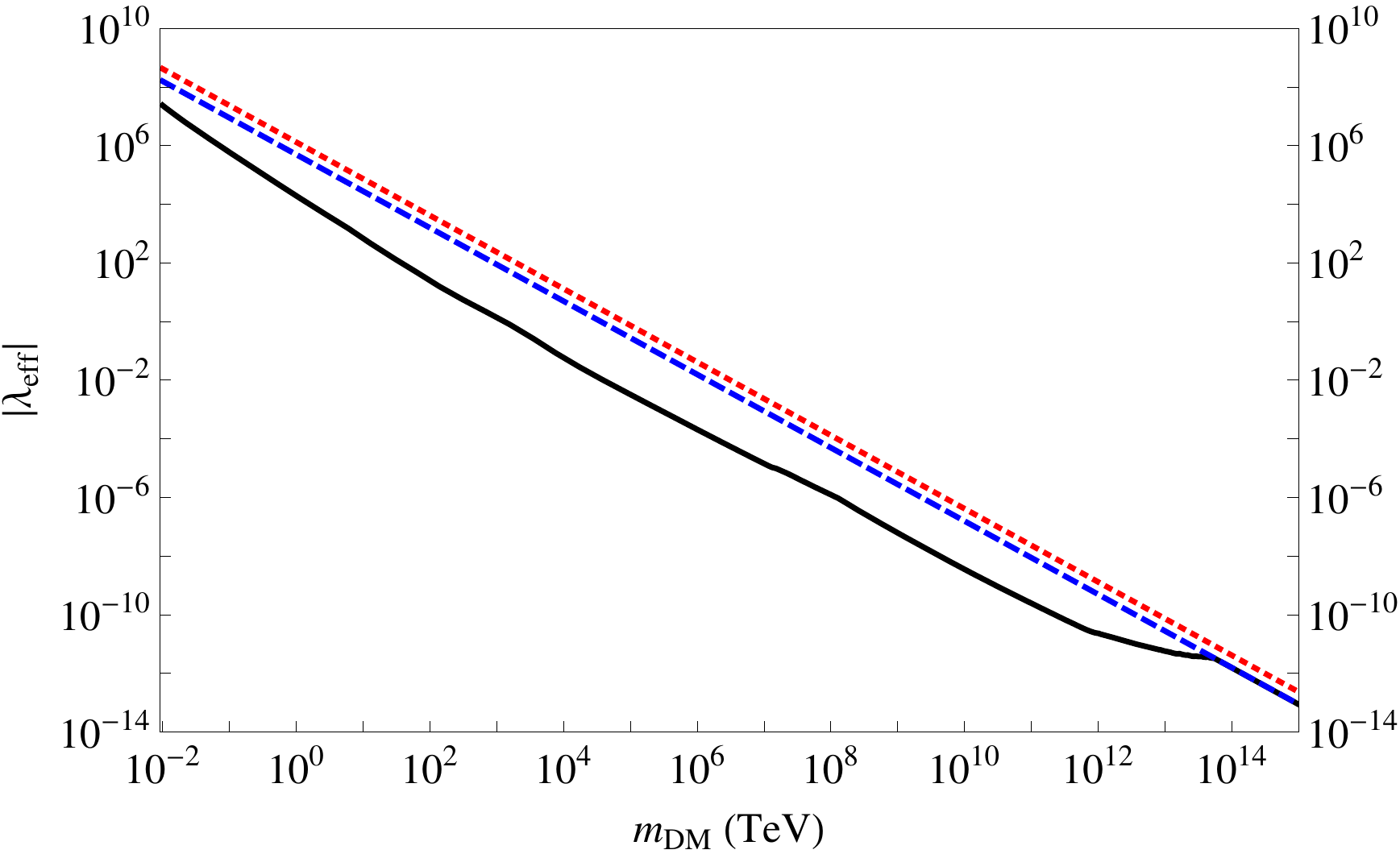}
\caption{Upper limit (90\% C. L.) on the parameter $\lambda_{\rm eff}$ which induces the three body decay $\psi\rightarrow e^+ e^- \nu$ (see Eq.~(\ref{eq:GammathreebodySigma})), assuming $M_\Sigma=M_P$. The dotted and dashed lines show respectively the limit from age of Universe and CMB. The solid line shows the limit obtained in this paper. }
\label{fig:lambdaeff3limit}
\end{center}
\end{figure}

Lastly, the width of the three body decay of a Majorana dark matter particle into an electron-positron pair and a neutrino $\psi_{\rm DM}\rightarrow e^+e^-\nu$ is fairly model dependent, as discussed in \cite{Garny:2010eg}. Let us first assume that the dark matter decay is mediated by a heavy charged scalar $\Sigma$. The Lagrangian reads:
\begin{align}
  \mathcal{L}_\text{eff}^\Sigma = -\bar{\psi}_{\rm DM} \left[\lambda_{e
  \psi}^L P_L + \lambda_{e \psi}^R P_R\right] e \, \Sigma^\dagger
  -\bar{\nu}\lambda_{e N}^R P_R  e
  \, \Sigma^\dagger + \text{h.c.}\;,
\end{align}
where $P_L = (1 - \gamma^5)/2$ and $P_R = (1 + \gamma^5)/2$ are the left- and
right-handed chirality projectors, respectively. Then, the dark matter decay width is given by:
\begin{equation}\label{eq:GammathreebodySigma}
\Gamma(\psi_{\rm DM}\rightarrow e^+e^-\nu)=
\frac{|\lambda_{\rm eff}|^4}{3072\pi^3}\frac{m_{\rm DM}^5}{M_\Sigma^4}
\end{equation}
where $|\lambda_{\rm eff}|^4=(|\lambda_{e \psi}^R|^2+|\lambda_{e \psi}^L|^2) |\lambda_{e \nu}^R|^2$. Fig.~\ref{fig:lambdaeff3limit} shows the limit on $\lambda_{\rm eff}$ as function of $m_{\rm DM}$ assuming $M_\Sigma=M_P$; when $M_\Sigma<M_P$ the limits scale as $(M_\Sigma/M_P)^4$.

In contrast, if the decay is mediated by a charged vector, the effective Lagrangian reads\footnote{We assume here that the decay is dominated by the charged-current interaction; in more generality the decay could also be mediated by a neutral current interaction.}
\begin{equation}
  \mathcal{L}_\text{eff}^V = -\bar{\psi}_{\rm DM} \gamma^\mu
  \left[\lambda_{e \psi}^L P_L + \lambda_{e \psi}^R P_R\right] e \,
  V_\mu^\dagger - \bar{\nu} \gamma^\mu 
  \lambda_{e N}^R P_R  e \, V_\mu^\dagger + \text{h.c.}\;.
\end{equation}
In this case the decay width is:
\begin{equation}\label{eq:GammathreebodyV}
\Gamma(\psi_{\rm DM}\rightarrow e^+e^-\nu)=
\frac{|\lambda_{\rm eff}|^4}{768\pi^3}\frac{m_{\rm DM}^5}{M_V^4}
\end{equation}
where, again, $|\lambda_{\rm eff}|^4=(|\lambda_{e \psi}^R|^2+|\lambda_{e \psi}^L|^2) |\lambda_{e \nu}^R|^2$. Note that the decay width for the decay mediated by a vector is a factor of four larger than the one mediated by a scalar, assuming $M_V=M_\Sigma$ in Eqs.~(\ref{eq:GammathreebodySigma}) and (\ref{eq:GammathreebodyV}).

\section{\label{sec:conc}Conclusions}

We have derived lower bounds on the lifetime of dark matter particles with masses in the range $10\,{\rm TeV} \leq m_{\rm DM}\leq 10^{15}\,{\rm TeV}$ from the non-observation of high energy neutrinos  in the experiments AMANDA, IceCube, Auger and ANITA. We have analyzed two scenarios where the dark matter decay produces monoenergetic neutrinos, namely a scalar dark matter particle which decays into a neutrino and an antineutrino and a fermionic dark matter particle which decays into a neutrino and a photon, as well as a scenario where the dark matter decay produces a neutrino flux with a softer spectrum, concretely a fermionic dark matter particle which decays into a neutrino and an electron-positron pair. We have found that, for dark matter masses between $\sim 10$ TeV and the Grand Unification scale, the lifetime is constrained to be larger than $\mathcal{O}(10^{26}-10^{28})~{\rm s}$, which is nine to eleven orders of magnitude longer than the age of the Universe, and seven to nine orders of magnitude stronger than the limit from the cosmic microwave background. We have also translated the limits on the lifetime into limits on the parameters of the Lagrangian of our three scenarios. Especially for large masses, the limits on the couplings are very strong and hint to the necessity of a symmetry ensuring the stability of the dark matter particle. 

\section*{Acknowledgements}

A.I. would like to thank the Instituto de F\'isica Gleb Wataghin at UNICAMP for hospitality during the preparation of this work. The work of A.I. was partially supported by the DFG cluster of excellence ``Origin and Structure of the Universe.'' The authors thank support from FAPESP and O. L. G. P. thanks support from CAPES/Fulbright.

\end{document}